\documentclass[aps,pra,floatfix,twocolumn,superscriptaddress]{revtex4-1}

\usepackage{amsmath}
\usepackage{graphicx}
\usepackage{ulem}
\usepackage{bbm}
\usepackage{bm}
\usepackage{color}
\usepackage{mathrsfs}
\usepackage{hyperref}

\hypersetup{backref,colorlinks=true,breaklinks,urlcolor=blue,linkcolor=blue,citecolor=blue}

\begin{document}

\title{Bosonic Weyl excitations induced by $p$-orbital interactions in a cubic optical lattice}

\author{Guang-Quan Luo}
\affiliation{Department of Physics, Southern University of Science and Technology, Shenzhen 518055, China}

\author{Guan-Hua Huang}
\affiliation{Department of Physics, Southern University of Science and Technology, Shenzhen 518055, China}

\author{Zhi-Fang Xu}
\email{xuzf@sustech.edu.cn}
\affiliation{Department of Physics, Southern University of Science and Technology, Shenzhen 518055, China}
\affiliation{Shenzhen Institute for Quantum Science and Engineering, Southern University of Science and Technology, Shenzhen 518055, China}

\begin{abstract}
Weyl points exist in a fascinating topological state of matter with linear band crossings analogous to magnetic monopoles. Tremendous efforts have been devoted to investigate fermionic topological matters with Weyl points in the single-particle band dispersion. It remains elusive for realizing interaction-induced Weyl points, especially for bosons. Motivated by recent experimental progress in ultracold atoms, we propose a scheme to create Weyl points for Bogoliubov excitations of a bosonic superfluid in a three-dimensional cubic optical lattice. The unique design of the lattice leads to interaction-induced time-reversal symmetry breaking for a $p$-orbital superfluid, which in turn induces Weyl Bogoliubov excitations. Analogous to Weyl semimetals of electronic systems, the superfluid also support topologically protected edge modes due to the bulk-boundary correspondence.
\end{abstract}

\maketitle
\section{Introduction}
\label{Introduction}
Weyl fermions, predicted by H. Weyl in 1929~\cite{weyl1929gravitation}, are massless spin-1/2 particles deriving from the Weyl equation in relativistic quantum field theory. They still haven't been directly observed as elementary particles. However, the existence of Weyl quasiparticles has been proved in real solid-state materials~\cite{PhysRevB.83.205101,PhysRevLett.107.127205,PhysRevX.5.011029,Huang2015,xu2015discovery,PhysRevX.5.031013,PhysRevX.5.031023}. Distinct from traditional topological insulators~\cite{PhysRevLett.61.2015,RevModPhys.82.3045,RevModPhys.83.1057,RevModPhys.88.021004}, the topological signature of Weyl materials is embodied in three-dimensional (3D) linear dispersion relation near Weyl points, which are paired band crossings and can not be removed unless the Weyl points with opposite charges are merged~\cite{turner2013beyond,wehling2014dirac,RevModPhys.90.015001}.

Ultracold atomic gases provide a highly controllable platform to realize desired quantum states of matter~\cite{Lewenstein2007,Bloch2012,Gross2017,Schafer2020}. Recently, a great number of significant progresses have been made for simulating Weyl points in ultracold atoms~\cite{PhysRevA.85.033640,PhysRevLett.114.225301,PhysRevA.92.013632,PhysRevA.94.013606,PhysRevA.94.043617,PhysRevA.95.033629,PhysRevResearch.1.033102,LU20202080,wang2021realization,PhysRevLett.112.136402,PhysRevLett.114.045302,PhysRevLett.115.265304,PhysRevB.96.035145,PhysRevA.104.033312,PhysRevA.95.023620}, especially an ideal Weyl semimetal band is realized in a quantum gas with 3D spin-orbit coupling~\cite{wang2021realization}. Most existing researches concentrate on Weyl points induced by spin-orbit couplings and artificial gauge fields in single-particle bands. A few studies focus on Weyl Bogoliubov excitations for fermionic gases~\cite{PhysRevLett.112.136402,PhysRevLett.114.045302,PhysRevLett.115.265304,PhysRevB.96.035145,PhysRevA.104.033312} and bosonic gases~\cite{PhysRevA.95.023620}. However, the relation between interactions and the emergence of Weyl points has remained elusive.

On the other hand, the investigation of interaction-induced topological phases for bosonic excitations is attracting increasing attention. In two-dimensional (2D) optical lattices, motivated by theoretical proposals~\cite{PhysRevLett.117.085301,PhysRevLett.117.163001,PhysRevA.98.053617,PhysRevA.103.043328}, interaction-induced atomic chiral superfluid with topological excitations in a Bose gas was realized in a recent experiment~\cite{Wang2021}. Even with these successful instances, the relevant researches are still lacking for the 3D topological phases. Similar to Weyl magnons~\cite{Li2016,PhysRevLett.117.157204,PhysRevB.95.224403,PhysRevB.95.085132,PhysRevB.96.104437,PhysRevB.97.094412,PhysRevB.97.115162}, interacting bosons in optical lattices also provide a new paradigmatic type of bosonic Weyl quasiparticle excitations.

In this paper, we propose a scheme to realize bosonic Weyl excitations induced by $p$-orbital interactions in a cubic optical lattice. The emergence of Weyl points needs either spatial inversion symmetry breaking or time-reversal symmetry breaking. The special repulsive interaction of $p$ orbitals can induce spontaneous time-reversal symmetry breaking for the bosonic superfluid~\cite{PhysRevA.74.013607}. We thus apply this mechanism to realize desired Weyl points for Bogoliubov excitations of the time-reversal symmetry breaking superfluid. The rest of the paper is organized as follows. In Sec.~\ref{Theoretical}, we introduce a cubic lattice by a layer-by-layer approach with a special design for $p$ orbitals. In Sec.~\ref{Bogoliubov}, we find the interaction will induce a $p_x\pm ip_y$ bosonic chiral superfluid with time-reversal symmetry breaking after loading bosons into $p$ orbitals. Therefore Weyl points, which are absent in the single-particle spectra, emerge in the Bogoliubov excitation spectra. In Sec.~\ref{Topology}, we discuss the effective two-band models near Weyl points which are obtained by the Krein-unitary perturbation theory deriving from the Schrieffer-Wolff transformation. The edge states of the bosonic excitation modes are also discussed. In Sec.~\ref{Experimental}, we propose a realistic scheme to create the cubic optical lattice in ultracold atom experiment. Distinct from the majority of previous proposals, our scheme does not rely on external Raman lasers or atomic internal states. This paper is concluded in Sec.~\ref{Conclusion}.

\section{Theoretical model}
\label{Theoretical}
\begin{figure}[tpb]
	\centering
	\includegraphics[width=8.6cm]{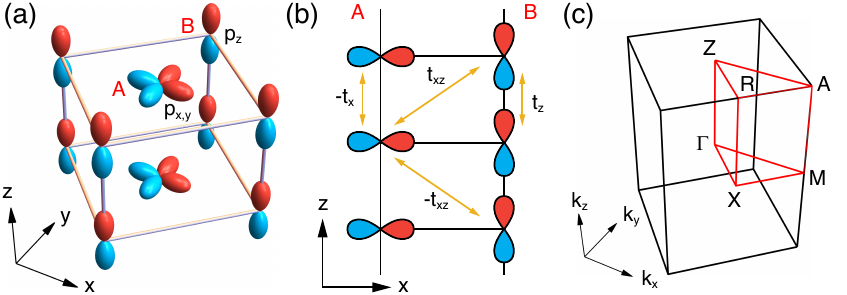}
	\caption{(a) The 3D cubic lattice comprises two classes of sites denoted by $A$ and $B$. (b) Illustration of the tight-binding tunnelings between $p$ orbitals on a 2D $xz$ plane. (c) The first Brillouin zone with the high-symmetry points.}
	\label{fig1}
\end{figure}

We construct a 3D cubic lattice via stacking a 2D square lattice along $z$ direction. One unit cell of the lattice contains two inequivalent lattice sites denoted by A and B, respectively. This is shown in Fig.~\ref{fig1}(a). Three $p$-orbitals with $p_{x,y}$ orbitals located at A sites and $p_z$ orbitals located at B sites are isolated from remaining orbitals by tuning the lattice parameters. We thus consider loading bosonic atoms into these three $p$-orbitals. The single-particle Hamiltonian can be written as
\begin{equation}
	\begin{aligned}
		\hat{H}_0=&\sum_{\mathbf{r}\in B,\pm}t_z\hat{p}^\dagger_{z,\mathbf{r}}\hat{p}_{z,\mathbf{r}\pm\mathbf{e}_z}+\sum_{\mathbf{r}\in B}\delta\hat{p}^\dagger_{z,\mathbf{r}}\hat{p}_{z,\mathbf{r}}\\
		&-\sum_{\mathbf{r}\in A,\pm}t_x\left(\hat{p}^\dagger_{x,\mathbf{r}}\hat{p}_{x,\mathbf{r}\pm\mathbf{e}_z}+\hat{p}^\dagger_{y,\mathbf{r}}\hat{p}_{y,\mathbf{r}\pm\mathbf{e}_z}\right)\\
		&+\sum_{\mathbf{r}\in A,m,n,\alpha}\left(mn t_{xz}\hat{p}^\dagger_{\alpha,\mathbf{r}}\hat{p}_{z,\mathbf{r}+m\mathbf{e}_\alpha+n\mathbf{e}_z}+\text{h.c.}\right).
		\label{tbm}
	\end{aligned}
\end{equation}
Here, $\hat{p}_{x,y}$ and $\hat{p}_z$ are bosonic annihilation operators for $p_{x,y}$ orbitals at A sites and $p_z$ orbitals at B sites, respectively. Tight-binding parameters $t_x$, $t_z$, $t_{xz}$, $\delta$ are all positive. $\mathbf{e}_{x,y,z}$ are vectors of length $a_0$ along $x,y,z$ directions, where $a_0$ is the distance between the nearest-neighbor sites. $m=\pm 1$, $n=\pm 1$, and $\alpha=x,y$. Tunnelings between $p_{x,y}$ and $p_z$ orbitals in the same $xy$ plane are prohibited due to odd parity of $p$-orbitals. Applying the Fourier transformation, the single-particle Hamiltonian in quasimomentum space can be obtained as $\hat{H}_0=\sum_\mathbf{k}\hat{\Psi}_\mathbf{k}^\dag \mathcal{H}_0( \mathbf{k})\hat{\Psi}_\mathbf{k}$, where $\hat{\Psi}_\mathbf{k}=(\hat{p}_{x,\mathbf{k}},\ \hat{p}_{y,\mathbf{k}},\ \hat{p}_{z,\mathbf{k}})^\mathrm{T}$ and $\mathcal{H}_0( \mathbf{k})$ is given by
\begin{equation}
\left( \begin{matrix}
-2t_x\cos(\tilde{k}_z)&		&    -4t_{xz}\sin(\tilde{k}_z)\sin(\tilde{k}_x)\\
&		-2t_x\cos(\tilde{k}_z)&    -4t_{xz}\sin(\tilde{k}_z)\sin(\tilde{k}_y)\\
\text{h.c.}&		&      \delta+2t_z\cos(\tilde{k}_z)\\
\end{matrix} \right),
\label{tbMatrix}
\end{equation}
where we have defined $\tilde{k}_{x,y,z}\equiv a_0k_{x,y,z}$.

The single-particle energy spectra are obtained by diagonalizing $\mathcal{H}_0( \mathbf{k})$. 
Since the system preserves time-reversal and spatial inversion symmetries, there is no Weyl point for the single-particle spectra. In the following parts, we consider to lift the $p_z$ orbitals comparing to $p_x$ and $p_y$ orbitals, which leads to a positive and large value of $\delta$. In this case, the energy minima are degenerate and located at the plane of $k_z=0$. This degeneracy can be broken if we consider more hopping terms, e.g. including next-to-next nearest neighbor tunnelings. For a real optical lattice system as discussed later, the energy minima are located at $M$ point and double degenerate (see Appendix~\ref{appendixC}). We thus consider a Bose-Einstein condensate with atoms condensing at $M$ point.

The interaction among atoms gives rise to the interaction-part Hamiltonian
\begin{equation}
\begin{aligned}
	\hat{H}_I =&\frac{U_{z}}{2}\sum_{\mathbf{r}\in B} \hat{p}_{z,\mathbf{r}}^\dagger\hat{p}_{z,\mathbf{r}}^\dagger\hat{p}_{z,\mathbf{r}}\hat{p}_{z,\mathbf{r}}\\
	&+ \frac{U_{x}}{2}\sum_{\mathbf{r}\in A}\left[ \hat{n}_{xy,\mathbf{r}}\left( \hat{n}_{xy,\mathbf{r}}-\frac{2}{3}\right)-\frac{\hat{L}^2_{z,\mathbf{r}}}{3}\right].\\
\end{aligned}
\end{equation}
Here, $\hat{n}_{xy,\mathbf{r}}=\hat{p}_{x,\mathbf{r}}^\dagger\hat{p}_{x,\mathbf{r}}+\hat{p}_{y,\mathbf{r}}^\dagger\hat{p}_{y,\mathbf{r}}$ and $\hat{L}_{z,\mathbf{r}}=-i(\hat{p}_{x,\mathbf{r}}^\dagger\hat{p}_{y,\mathbf{r}}-\hat{p}_{y,\mathbf{r}}^\dagger\hat{p}_{x,\mathbf{r}})$. $U_{z}$ and $U_{x}$ are positive for the repulsive interaction. Obviously, the interaction between atoms favors a ground state with $\langle \hat{L}_{z,\mathbf{r}} \rangle \neq 0$, leading to a time-reversal symmetry breaking condensate.

\section{Bogoliubov excitations}
\label{Bogoliubov}
\begin{figure}[htbp]
	\centering
	\includegraphics[width=8.6cm]{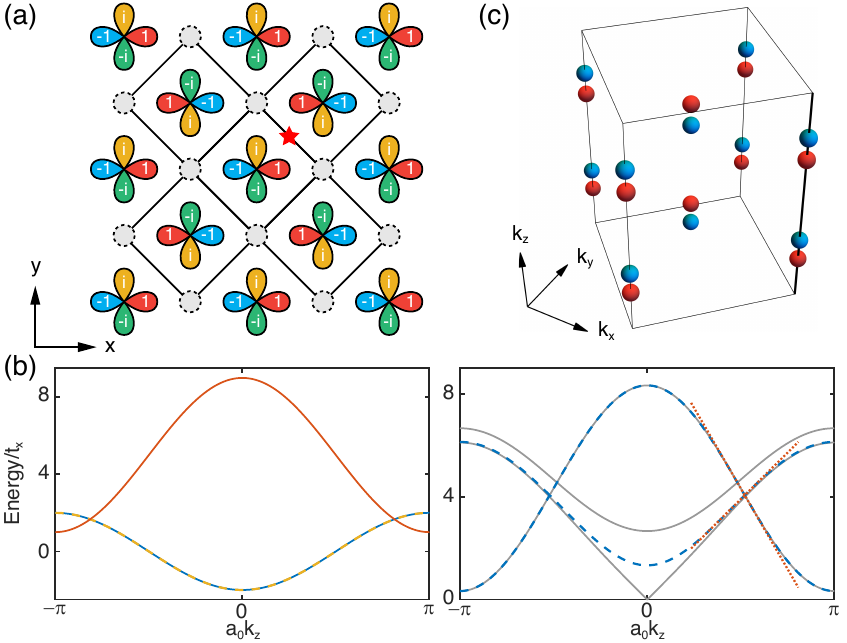}
	\caption{(a) Schematic diagram for the superfluid ground state of $\bm{\varphi}_+$ in the $xy$ slice. The gray circles with dashed edge indicate the positions of B sites. The red star is a center of the inversion symmetry. (b) The single-particle energy spectra (left panel) and the Bogoliubov excitation spectra (right panel). The spectra are shown along $(k_x,k_y)=(\pi/a_0,0)$, which is indicated by the bold black lateral edge in (c). The parameters are chosen as $t_z/t_x=2$, $t_{xz}/t_x=0.5$, $\delta/t_x=5$, $U_x\rho/t_x=4$, and $U_z\rho/t_x=4$. (c) The Weyl points in first Brillouin zone. The red and blue spheres denote the Weyl points with positive and negative charges, respectively.}
	\label{fig2}
\end{figure}

We use the mean-field theory to determine the ground state. Since the atoms are assumed to condense at $M$ point, we obtain the ground-state energy functional as
\begin{equation}
\begin{aligned}
E_\mathrm{MF}(\bm{\varphi})=&\bm{\varphi}^\dagger \mathcal{H}_0(\mathbf{M})\bm{\varphi}+\frac{U_z\rho}{2}|\varphi_3|^4+\frac{U_x\rho}{2}(|\varphi_1|^4+|\varphi_2|^4)\\
&+\frac{U_x\rho}{6}(4|\varphi_1|^2|\varphi_2|^2+\varphi_1^{*2}\varphi_2^2+\varphi_2^{*2}\varphi_1^2).
\end{aligned}
\end{equation}
Here, we define $\bm{\varphi}\equiv(1/\sqrt{N_0})(\langle\hat{p}_{x,\mathbf{M}}\rangle,\langle\hat{p}_{y,\mathbf{M}}\rangle,\langle\hat{p}_{z,\mathbf{M}}\rangle)^\mathrm{T}=(\varphi_1,\varphi_2,\varphi_3)^\mathrm{T}$, and $N_0$ is the particle number of condensate. $\rho$ is the atom number per unit cell. Via minimization of the energy functional, we obtain two degenerate ground states with $\bm{\varphi}=\bm{\varphi}_\pm$, where $\bm{\varphi}_\pm\equiv(1/\sqrt{2})(1,\pm i,0)^\mathrm{T}$. The corresponding order parameters are given by
\begin{equation}
\langle \hat{\Psi} (\mathbf{r}) \rangle=e^{i\mathbf{M}\cdot \mathbf{r}}
\frac{1}{\sqrt{2}}(1,\pm i, 0)^\mathrm{T}.
\end{equation}
Each ground state breaks the time-reversal symmetry and in the meantime preserves the inversion symmetry. One ground state is shown in Fig.~\ref{fig2}(a) for the case with weak interaction among atoms.

We then investigate the Bogoliubov excitations for the time-reversal symmetry breaking ground state with $\bm{\varphi}=\bm{\varphi}_+$. The Bogoliubov-de Gennes (BdG) Hamiltonian can be written as
$\hat{H}_{\mathrm{BdG}}=(1/2)\sum_{\mathbf{k}\neq\mathbf{M}}
\hat{\bm{\phi}}_\mathbf{k}^\dagger
\mathcal{H}_{\mathrm{BdG}}(\mathbf{k})
\hat{\bm{\phi}}_\mathbf{k}$, where $\hat{\bm{\phi}}_\mathbf{k}=(\hat{p}_{x,\mathbf{k}},\hat{p}_{y,\mathbf{k}},\hat{p}_{z,\mathbf{k}},\hat{p}_{x,2\mathbf{M}-\mathbf{k}}^\dagger\hat{p}_{y,2\mathbf{M}-\mathbf{k}}^\dagger,\hat{p}_{z,2\mathbf{M}-\mathbf{k}}^\dagger)$ and $\mathcal{H}_{\mathrm{BdG}}(\mathbf{k})$ takes the form of
\begin{equation}
\mathcal{H}_{\mathrm{BdG}}(\mathbf{k})=\left(\begin{array}{cc}
h_{a}(\mathbf{k}) & h_{x}(\mathbf{k})\\
h_{x}^{*}(-\mathbf{k}) & h_a^*(-\mathbf{k})
\end{array}\right).
\end{equation}
Note that we use ``$-\mathbf{k}$'' rather than ``$2\mathbf{M}-\mathbf{k}$'' because the quasimomentum $2\mathbf{M}$ is equivalent to zero. The matrix $h_a(\mathbf{k})$ is given by $h_a(\mathbf{k})=\mathcal{H}_0(\mathbf{k})-\mu 1_{3\times 3}+\rho \mathrm{diag}(4U_x/3,4U_x/3,0)$. Here $\mu=E_0(\mathbf{M})+2U_x\rho/3$ with $E_0(\mathbf{M})=-2t_x$ being the single-particle energy of ground state, $1_{3\times 3}$ is the identity matrix. The matrix $h_{x}(\mathbf{k})$ is given by
\begin{equation}
h_{x}(\mathbf{k})=\frac{U_x\rho}{3}\left(\begin{array}{ccc}
		1 & i & 0\\
		i & -1 & 0\\
		0 & 0 & 0
	\end{array}\right).
\end{equation}

To satisfy the bosonic commutation relation, the bosonic BdG Hamiltonian $\mathcal{H}_{\mathrm{BdG}}(\mathbf{k})$ is diagonalized by a paraunitary matrix $T_\mathbf{k}$ rather than a unitary matrix as $T_\mathbf{k}^\dagger \mathcal{H}_{\mathrm{BdG}}(\mathbf{k}) T_\mathbf{k} = \mathrm{diag} (\mathcal{E}_\mathbf{k}, \mathcal{E}_{-\mathbf{k}})$. The paraunitary matrix $T_\mathbf{k}$ satisfies the relations of $T_\mathbf{k}^\dagger\tau_zT_\mathbf{k}=\tau_z$ and $T_\mathbf{k}\tau_zT_\mathbf{k}^\dagger=\tau_z$, where $\tau_z=\sigma_z \otimes 1_{3\times 3}$ with $\sigma_z$ being the Pauli matrix. The excitation spectra $\mathcal{E}_\mathbf{k}$ along the line with $(k_x,k_y)=(\pi/a_0,0)$ are depicted with the gray solid lines in the right panel of Fig.~\ref{fig2}(b). We note that the degeneracy among $p_{x,y}$ bands (marked by the blue solid line and the yellow dashed line in the left panel of Fig.~\ref{fig2}(b)) is lifted by the atomic interaction for the Bogoliubov excitations. The corresponding excitation spectra are given by
\begin{eqnarray}
	\mathcal{E}_1(k_z)&=&2|\sin\frac{\tilde{k}_{z}}{2}|\sqrt{2t_{x}\left(t_{x}+\frac{2U_{x}\rho}{3}-t_{x}\cos \tilde{k}_{z}\right)},\nonumber\\
	\mathcal{E}_2(k_z)&=&2t_{x}+\frac{2U_{x}\rho}{3}-2t_{x}\cos \tilde{k}_{z},\nonumber\\
  \mathcal{E}_3(k_z)&=&2t_{x}-\frac{2U_{x}\rho}{3}+2t_{z}\cos \tilde{k}_{z}+\delta,
\end{eqnarray}
where $\mathcal{E}_i$ denotes the excitation energy for the $i$-th excitation band. These feature also appears for excitation spectra along the line with $(k_x,k_y)=(0,0)$.

Therefore, numerous band crossing points emerge for the Bogoliubov excitations of the time-reversal symmetry breaking superfluid. As we confirmed in the following section, these crossing points are indeed Weyl points. In total, we find that  there are eight Weyl points. Half of them are located at the line with $(k_x,k_y)=(0,0)$ and the other half are located at the line with $(k_x,k_y)=(\pi/a_0,0)$.  On each line, two Weyl points arise at the band crossing point between the first and second excitation bands (characterized by $\mathcal{E}_1$ and $\mathcal{E}_2$) and another two Weyl points are located at band crossing points between the second and third excitation bands (characterized by $\mathcal{E}_2$ and $\mathcal{E}_3$). The topological charges of the Weyl points can be obtained by numerical calculations. We show the positions and charges of the Weyl points in Fig.~\ref{fig2}(c).

\section{Topology and edge states}
\label{Topology}
A Hamiltonian to describe the dispersion for a Weyl point can be written as $\mathcal{H}_W(\mathbf{k})=\sum_{i,j}k_iv_{ij}\sigma_j$, where $j=x,y,z$. The topological charge of a Weyl point is determined by $c=\text{sgn}[\det (v_{ij})]$, which equals to $\pm 1$. We apply the perturbation transformation on the bosonic BdG Hamiltonian $\mathcal{H}_{\mathrm{BdG}}(\mathbf{k})$ to obtain an effective low-energy Hamiltonian to describe Weyl points. More details can be found in Appendix~\ref{appendixA} and the related literature~\cite{Zhou_2020,PhysRevA.103.013308,PhysRevB.106.144434}.

We first define a non-Hermitian matrix $H\equiv\tau_zH_\mathrm{BdG}(\mathbf{k})$, which is Krein Hermitian with respect to $\tau_z$. $H$ can be decomposed into a diagonal matrix $H^0$ and a perturbation part $H^\prime$, where $H^\prime$ can be further decomposed into a block-diagonal matrix $H^1$ and a block off-diagonal matrix $H^2$. We can block-diagonalize $H$ by the Schrieffer-Wolff transformation
\begin{equation}
	\begin{aligned}
		\tilde{H}&=e^{-W}He^{W}\\
		&=H+[H,W]+\frac{1}{2}[[H,W],W]+\cdots,
	\end{aligned}
\end{equation}
where $e^W$ is Krein-unitary. Up to second order, $\tilde{H}$ has the form
\begin{equation}
\tilde{H}=H^0+H^1+\frac{1}{2}[H^2,W],
\end{equation}
where $W$ is determined by $[H^0,W]+H^2=0$.

The Weyl points emerge at the linear band crossings and are mainly induced by the hybridization  among $p_+$ and $p_z$ bands or $p_-$ and $p_z$ bands in the excitation spectra, where $p_\pm =p_x \pm i p_y$. Therefore, we should transform the bases from $p_{x,y}$ orbitals to $p_{\pm}$ orbitals, and then only focus on the subspace of $p_{+,z}$ or $p_{-,z}$ during the perturbation process. The perturbation method is not available for the entire Brillouin zone due to the divergence, but it does not influence the series expansion near the Weyl points. We apply the series expansion up to first order obtain the final result which satisfies the form of the Weyl Hamiltonian $\mathcal{H}_W(\mathbf{k})$. We thus can derive topological charges for Weyl points. In the right panel of Fig.~\ref{fig2}(b), the energy spectra for the effective model with only $p_{+,z}$ orbitals and the series expansion near a Weyl point in the effective model are depicted with the blue dashed lines and the red dashed lines, respectively.

The topological charge of a Weyl point is also associated with the Chern number of a 2D surface surrounding the Weyl point. Based on the discussion in Ref.~\cite{PhysRevB.87.174427}, the Chern number of $n$-th band can be obtained by the integral of Berry curvature as $c_n=1/(2\pi)\int d\mathbf{k} \mathbf{B}_n(\mathbf{k})$. The Berry curvature is defined as $\mathbf{B}_n(\mathbf{k})=\nabla_\mathbf{k}\times \mathbf{A}_n(\mathbf{k})$ with $\mathbf{A}_n(\mathbf{k})=i\langle T^n_\mathbf{k}|\tau_z\nabla_\mathbf{k}|T^n_\mathbf{k}\rangle$, where $|T^n_\mathbf{k}\rangle$ is the $n$-th column eigenvector of $T_\mathbf{k}$. Numerically, we consider all six surfaces of a tiny cube enclosing a Weyl point and calculate their Berry curvatures by the algorithm described in Ref.~\cite{fukui2005chern}. The Chern number of all surfaces is the topological charge of the Weyl point. The resulting Chern numbers exactly match with the charges obtained from the series expansion. We show the charge of each Weyl point in Fig.~\ref{fig2}(c) with $c = + 1$ and $c = - 1$ being distinguished by red and blue colors, respectively.

\begin{figure}[htbp]
	\centering
	\includegraphics[width=8.6cm]{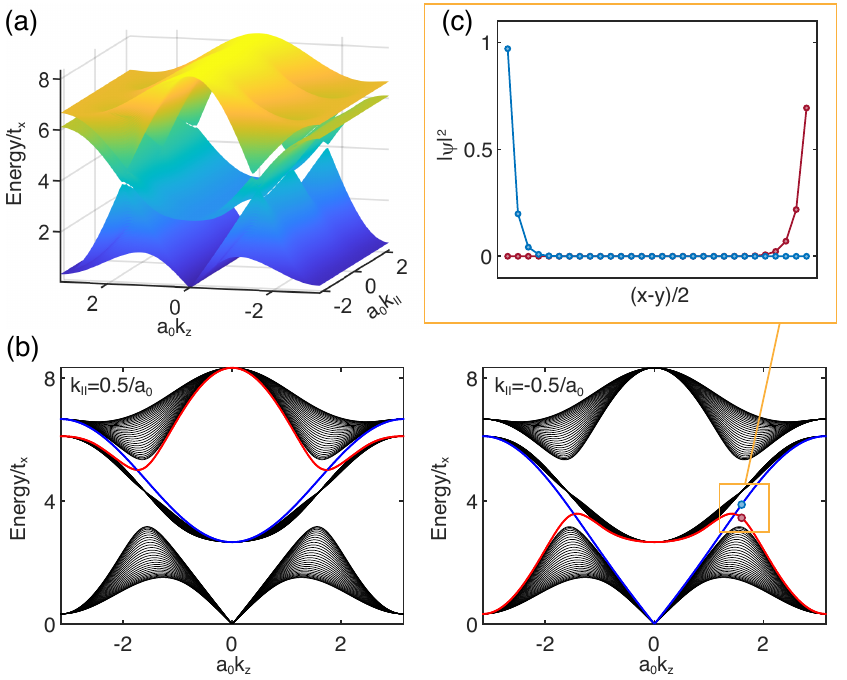}
	\caption{(a) The Bogoliubov excitation spectra for the semi-infinite system of the 3D cubic lattice. (b) The slices of the excitation spectra at $k_\parallel=\pm 0.5/a_0$ in (a). (c) The typical edge states in (b). The right and left edge states are indicated by red and blue colors. The parameters are chosen as used in Fig.~\ref{fig2}.}
	\label{fig3}
\end{figure}

Due to bulk-edge correspondence, there are also topological protected edge states analogous to the Fermi arcs in electronic systems for the open system. We thus set an open boundary along $(\hat{x}-\hat{y})/\sqrt{2}$ direction and a periodic boundary along $(\hat{x}+\hat{y})/\sqrt{2}$ and $\hat{z}$ directions, where $k_\parallel$ denotes the quasimomentum component along $(\hat{x}+\hat{y})/\sqrt{2}$. One unit cell of the semi-infinite systems contains 30 unit cells of the lattice. In Fig.~\ref{fig3}(a) and (b), we show the energy spectra of the semi-infinite system. The density distributions of typical edge states on the lattice surfaces illustrate the bulk-boundary correspondence as shown in Fig.~\ref{fig3}(c). More details can be found in Appendix~\ref{appendixB}.

\section{Experimental implementation}
\label{Experimental}
A 2D bipartite square optical lattice has been realized in the experiment~\cite{Wirth2011}. This optical lattice comprises a deep potential well and a shallow potential well in one unit cell. The lattice potential in $xy$ plane is given by
\begin{equation}
V_\mathrm{2D}(x,y)=-V_0\left|\cos(k_Lx)+\exp(i\theta)\cos(k_Ly)\right|^2.
\end{equation}
Here, $k_L=2\pi/\lambda$ with $\lambda=1064$ nm being the wavelength of laser beams. This bipartite square lattice is the 2D counterpart of the 3D cubic optical lattice we desired.

\begin{figure}[htbp]
	\centering
	\includegraphics[width=8.6cm]{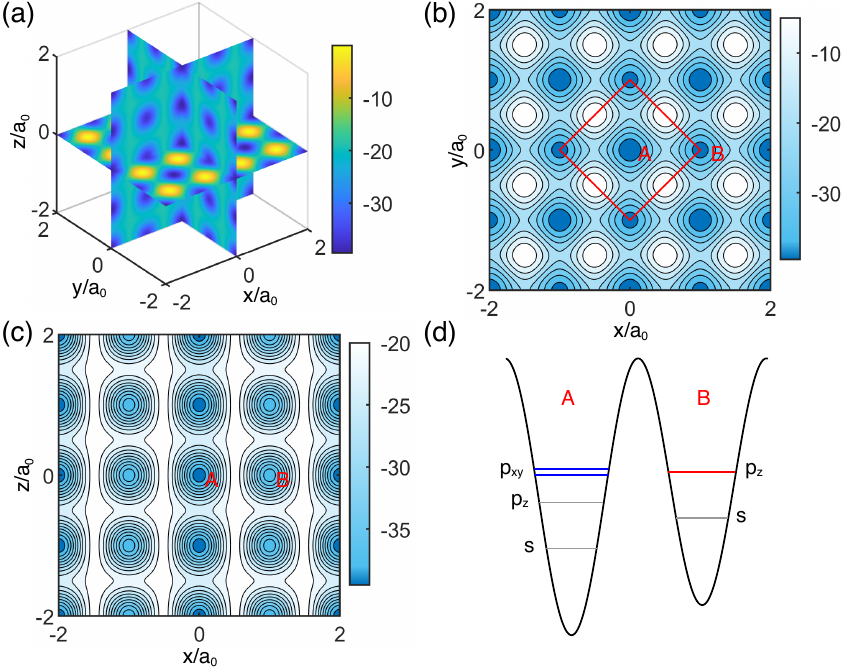}
	\caption{(a) The 3D cubic optical lattice potential $V(\mathbf{r})$ with $(V_0,\theta,V^\prime)=(11 E_\mathrm{R},0.484\pi,4.1 E_\mathrm{R})$, where $E_\mathrm{R}=\hbar^2k_L^2/(2m_a)$ with $m_a$ being the mass of the particle. (b) The contour plot of the potential slice at $z=0$. (c) The contour plot of the potential slice at $y=0$. (d) Schematic picture for the energy difference between $p_{x,y}$ orbitals at A site, $p_z$ orbital at B site and other orbitals.}
	\label{fig4}
\end{figure}

We further consider to add extra beams propagating along $x\pm z$ and $y\pm z$ directions to create a 3D cubic lattice with lattice potential  $V(\mathbf{r})=V_\mathrm{2D}(x,y)+V^{\prime}(\mathbf{r})$, where 
\begin{eqnarray}
V^{\prime}(\mathbf{r})=-V^\prime\cos^{2}[k_{L}(x+z)]-V^\prime\cos^{2}[k_{L}(x-z)]\nonumber\\
-V^\prime\cos^{2}[k_{L}(y+z)]-V^\prime\cos^{2}[k_{L}(y-z)],
\end{eqnarray}
As illustrated in Fig.~\ref{fig4}, the 3D optical lattice comprises two potential wells in one unit cell and the relative energy offset between the two wells is adjustable. We label the deep well as A and the shallow well as B. The horizontal and vertical slices of the lattice potential are shown in Fig.~\ref{fig4}(b) and (c). The unit cell is a cuboid and the nearest distance between A and B is $a_0=\pi/k_L$. By choosing suitable potential parameters, $p_{x,y}$ orbitals in A site and $p_z$ orbital in B site are coupled and far away from other orbitals as shown in Fig.~\ref{fig4}(d).

We also perform plane-wave expansion to calculation the energy spectra for the 3D lattice potential and confirm that the tight-binding model of Eq.~(\ref{tbm}) indeed well describe the band structure. A more practical tight-binding model with more hoppings and the corresponding Bogoliubov excitations have also been discussed. These can be found in Appendix~\ref{appendixC}. 

One may load the atoms into the $p$ orbital bands by applying the band swapping method. The interaction strength can be adjusted by the Feshbach resonance technique. The Weyl excitations of the bosonic superfluid may be detected by the momentum-resolved Bragg spectroscopy~\cite{Ernst2010}. A density wave can form along the edge owing to an interference of the background condensate and the edge mode~\cite{Furukawa_2015}. The Hall responses of the bosonic topological Bogoliubov excitations have also been discussed in Ref.~\cite{PhysRevLett.129.185301}.

\section{Conclusion}
\label{Conclusion}
We investigate bosonic Weyl excitations in a 3D cubic optical lattice with Weyl points emerging in the Bogoliubov excitation spectra. The bosonic Weyl excitations are induced by the interaction of $p$ orbitals. Furthermore, an possible experimental scheme based on the existing experiment of a 2D bipartite optical square lattice is also proposed. This work complements the search of Weyl systems in orbital optical lattices and motivates future experimental studies with realistic ultracold atomic platforms.

\begin{acknowledgments}
We acknowledge the helpful discussion with Liang-Liang Wan. This work is supported by the National Key R\&D Program of China (Grants No.~2022YFA1404103 and No.~2018YFA0307200),  NSFC (Grant No.~12274196), and a fund from Guangdong province (Grant No.~2019ZT08X324).
\end{acknowledgments}

\appendix
\section{Effective Hamiltonian near Weyl points}
\label{appendixA}
A general anisotropic Weyl Hamiltonian can be written as $\mathcal{H}_W(\mathbf{k})=A_0+\sum_iA_ik_i\sigma_0+\sum_{i,j}k_iv_{ij}\sigma_j$, where $\sigma_j$ are Pauli matrices and $i,j=x,y,z$. To confirm the Weyl points and their charges in the excitation spectra, we try to obtain the low-energy $2\times 2$ Hamiltonians near the Weyl points from the original $6\times 6$ bosonic BdG Hamiltonian $\mathcal{H}_\mathrm{BdG}(\mathbf{k})$, the final results are expected to be in the form of $\mathcal{H}_W(\mathbf{k})$. Due to the constraint of bosonic commutation relation, the eigenmodes of a $2n\times 2n$ Hermitian matrix $H_\mathrm{BdG}(\mathbf{k})$ are solved by a congruence transformation $T_\mathbf{k}^\dagger H_\mathrm{BdG}(\mathbf{k})T_\mathbf{k}$. The transformation matrix $T_\mathbf{k}$ is paraunitary and satisfies
\begin{equation}
	T^\dagger_\mathbf{k} \tau_zT_\mathbf{k}=T_\mathbf{k}\tau_zT^\dagger_\mathbf{k}=\tau_z
\end{equation}
with $\tau_z=\sigma_z\otimes 1_{n\times n}$. The congruence transformation can be rewritten as a similarity transformation of a non-Hermitian matrix $\tau_z H_\mathrm{BdG}(\mathbf{k})$
\begin{equation}
	T_\mathbf{k}^\dagger H_\mathrm{BdG}(\mathbf{k})T_\mathbf{k}=\tau_z (T_\mathbf{k}^{-1}(\tau_z H_\mathrm{BdG}(\mathbf{k}))T_\mathbf{k}).
\end{equation}
The matrix $\tau_z H_\mathrm{BdG}(\mathbf{k})$ and the transformation $T_\mathbf{k}$ are Krein-Hermitian and Krein-unitary with respect to $\tau_z$, respectively. In the following contents of this section, we define $H\equiv\tau_zH_\mathrm{BdG}(\mathbf{k})$.

We then apply the perturbation theory to approximately describe Weyl points. We decompose $H$ into a zeroth order diagonal matrix $H^0$ and a first order perturbation $H^\prime$
\begin{equation}
	H=H^0+H^\prime=H^0+H^1+H^2,
\end{equation}
where $H^1$ is block-diagonal with two subsets and $H^2$ is block off-diagonal. $H$ can be block-diagonalized by the Schrieffer-Wolff transformation
\begin{equation}
	\begin{aligned}
		\tilde{H}&=e^{-W}He^{W}\\
		&=H+[H,W]+\frac{1}{2}[[H,W],W]+\cdots,
	\end{aligned}
\end{equation}
where $e^W$ is a paraunitary matrix and $W$ is block off-diagonal like $H^2$. Consider the first order of the block off-diagonal part of $\tilde{H}$ is zero, we find
\begin{equation}
	[H^0,W]+H^2=0,
\end{equation}
where $W$ only up to first order can be obtained by
\begin{equation}
	W_{ml}=-\frac{H_{ml}^\prime}{E_{m}^0-E_{l}^0},
	\label{EqW}
\end{equation}
where $E_{m}^0$ are the diagonal elements of $H^0$. Up to second order, $\tilde{H}$ is given by
\begin{equation}
	\tilde{H}=H^0+H^1+\frac{1}{2}[H^2,W]+\cdots.
\end{equation}
The matrix elements of $\tilde{H}$ can be expressed as
\begin{equation}
\begin{aligned}
\tilde{H}_{mm^\prime}=&H_{mm^{\prime}}^0+H_{mm^{\prime}}^\prime\\
&+\frac{1}{2}\sum_{l}H_{ml}^\prime H_{lm^{\prime}}^\prime \left(\frac{1}{E_{m}^0-E_{l}^0}+\frac{1}{E_{m^{\prime}}^0-E_{l}^0}\right).
\end{aligned}
\label{Eq2rd}
\end{equation}
This result is obtained by taking up to second order perturbation. In fact, this perturbation method can give rise to the effective model in an arbitrary order~\cite{PhysRevB.106.144434}.

\begin{figure}[tbp]
	\centering
	\includegraphics[width=8.6cm]{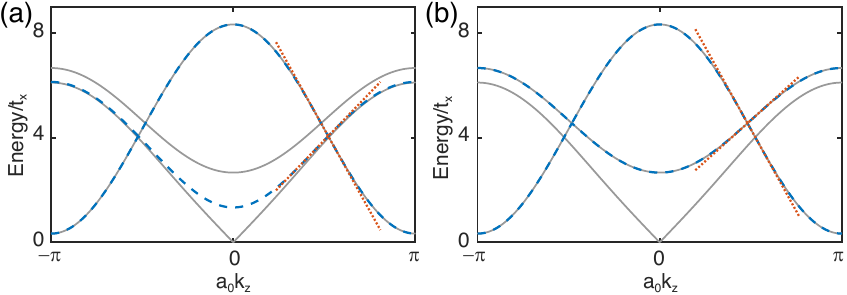}
	\caption{The gray lines indicate the original excitation spectra. The blue dashed lines indicate the energy bands from (a) the effective model of $p_+$ and $p_z$ orbitals and (b) the effective model of $p_-$ and $p_z$ orbitals. The red dashed lines indicate the results based on series expansions near the Weyl points in the effective models. The energy spectra are shown along the line of $(k_x,k_y)=(\pi/a_0,0)$. The parameters are the same as that used in Fig.~\ref{fig2} of the main text.}
	\label{sfig1}
\end{figure}
By defining $\hat{p}_{\pm}^\dagger=\hat{p}_x^\dagger\pm i \hat{p}_y^\dagger$, we can rewrite the bosonic BdG Hamiltonian via
\begin{equation}
	\hat{\bm{\phi}}_\mathbf{k}^\dagger
	\mathcal{H}_{\mathrm{BdG}}(\mathbf{k})
	\hat{\bm{\phi}}_\mathbf{k}=\hat{\bm{\phi}}_\mathbf{k}^{\prime\dagger}
	U^\dagger \mathcal{H}_{\mathrm{BdG}} (\mathbf{k}) U
	\hat{\bm{\phi}}_\mathbf{k}^\prime,
\end{equation}
where
$\hat{\bm{\phi}}_\mathbf{k}^\prime=(\hat{p}_{+,\mathbf{k}},\hat{p}_{-,\mathbf{k}},\hat{p}_{z,\mathbf{k}},\hat{p}_{+,2\mathbf{M}-\mathbf{k}}^\dagger\hat{p}_{-,2\mathbf{M}-\mathbf{k}}^\dagger,\hat{p}_{z,2\mathbf{M}-\mathbf{k}}^\dagger)$ and
\begin{equation}
	U=\left(\begin{array}{ccc}
		1 & 1\\
		i & -i\\
		&  & \sqrt{2}
	\end{array}\right)\oplus\left(\begin{array}{ccc}
		1 & 1\\
		-i & i\\
		&  & \sqrt{2}
	\end{array}\right)/\sqrt{2}.
\end{equation}
Since Weyl points emerge at the linear band crossings and are mainly induced by the hybridization  among $p_+$ and $p_z$ bands or $p_-$ and $p_z$ bands in the excitation spectrum. Therefore, we should transform the bases from $p_{x,y}$ orbitals to $p_{\pm}$ orbitals, and then only focus on the subspace of $p_{+,z}$ or $p_{-,z}$ during the perturbation process. The second order perturbation method is applied for $H=\tau_z(U^\dagger\mathcal{H}_{\mathrm{BdG}} (\mathbf{k}) U)$. The energy spectra of the effective two-band models are shown in Fig~\ref{sfig1}. We further obtain all effective $\mathbf{k}\cdot \mathbf{p}$ Hamiltonians near Weyl points up to first order of $\mathbf{k}$ and find that the topological charges from $c=\text{sgn}[\det (v_{ij})]$ are consistent with the numerically calculated Chern numbers. For a typical Weyl Hamiltonian near the crossing point at $(k_x,k_y,k_z)=(\pi/a_0,0,1.639/a_0)$ due to the hybridization among $p_+$ and $p_z$ orbitals, we find $A_0=4.0620$, $A_{z}=-0.8439a_0$, $v_{xx}=1.4110a_0$, $v_{yy}=-1.4110a_0$, $v_{zz}=3.1469a_0$, and other elements of $A_i$ and $v_{ij}$ are zero. For another typical Weyl Hamiltonian near the crossing point of $(k_x,k_y,k_z)=(\pi/a_0,0,1.1512/a_0)$ due to the hybridization among $p_-$ and $p_z$ orbitals, we find $A_0=4.5556$, $A_{z}=-0.9985a_0$, $v_{xx}=1.4120a_0$, $v_{yy}=1.4120a_0$, $v_{zz}=2.9954a_0$, and other elements of $A_i$ and $v_{ij}$ are zero.

\section{Edge state densities of the bosonic excitation modes}
\label{appendixB}
\begin{figure}[htbp]
	\centering
	\includegraphics[width=8.6cm]{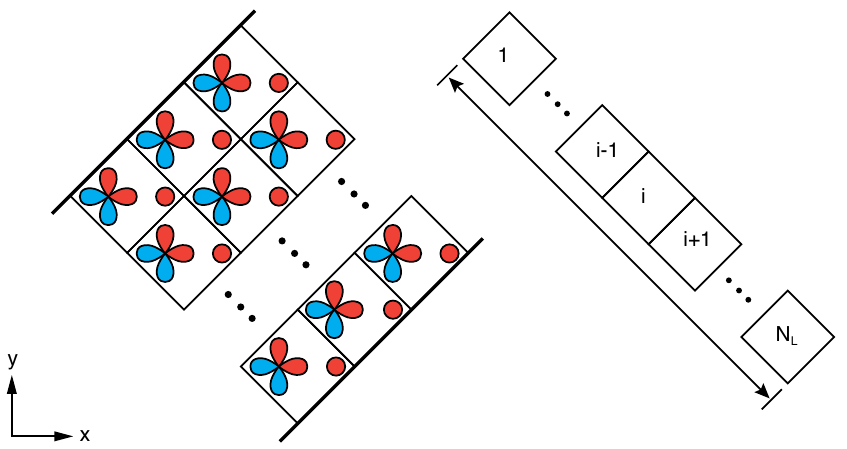}
	\caption{Schematic diagram of the semi-infinite lattice structure. The bold black lines indicate the open boundaries. An extended unit cell include $N_L$ original unit cells and each original unit cell includes three orbitals of $p_x$, $p_y$, and $p_z$.}
	\label{sfig2}
\end{figure}
We consider a semi-infinite lattice structure as shown in the left panel of Fig.~\ref{sfig2}. The periodic boundaries are applied along $(\hat{x}+\hat{y})/\sqrt{2}$ and $\hat{z}$ directions, their corresponding quasimomenta are denoted by $k_\parallel$ and $k_z$. Open boundary condition is applied along the $(\hat{x}-\hat{y})/\sqrt{2}$ direction. One unit cell of the semi-infinite lattice contains $N_L$ unit cells of the original lattice as shown in the right panel of Fig.~\ref{sfig2}. Consider that the quasimomentum of the condensate is $\mathbf{M}$, the bosonic BdG Hamiltonian is diagonalized as
\begin{equation}
	\begin{aligned}
		\hat{H}_{\mathrm{BdG}}&=\frac{1}{2}\sum_{\mathbf{k}\neq \mathbf{M}}
		\left(\hat{\boldsymbol{\beta}}_{\mathbf{k}}^{\dagger},\hat{\boldsymbol{\beta}}_{2\mathbf{M}-\mathbf{k}}^\top\right)
		\mathcal{H}_{\mathrm{BdG}}(\mathbf{k})
		\left(\begin{array}{c}
			\hat{\boldsymbol{\beta}}_{\mathbf{k}}\\
			\hat{\boldsymbol{\beta}}_{2\mathbf{M}-\mathbf{k}}^{\dagger\top}
		\end{array}\right)\\
		&=\frac{1}{2}\sum_{\mathbf{k}\neq \mathbf{M}}
		\left(\hat{\boldsymbol{\gamma}}_{\mathbf{k}}^{\dagger},\hat{\boldsymbol{\gamma}}_{2\mathbf{M}-\mathbf{k}}^\top\right)
		T_\mathbf{k}^\dagger\mathcal{H}_{\mathrm{BdG}}(\mathbf{k})T_\mathbf{k}
		\left(\begin{array}{c}
			\hat{\boldsymbol{\gamma}}_{\mathbf{k}}\\
			\hat{\boldsymbol{\gamma}}_{2\mathbf{M}-\mathbf{k}}^{\dagger\top}
		\end{array}\right).
	\end{aligned}
\end{equation}
Here, $\hat{\boldsymbol{\beta}}_{\mathbf{k}}=(\hat{\beta}_{1,\mathbf{k}},\hat{\beta}_{2,\mathbf{k}},\cdots)^\mathrm{T}$ are the bosonic annihilation operators. $T_\mathbf{k}^\dagger\mathcal{H}_{\mathrm{BdG}}(\mathbf{k})T_\mathbf{k}$ is diagonal and $\hat{\boldsymbol{\gamma}}_{\mathbf{k}}=(\hat{\gamma}_{1,\mathbf{k}},\hat{\gamma}_{2,\mathbf{k}},\cdots)^\mathrm{T}$ are the annihilation operators of the Bogoliubov excitation modes. The paraunitary matrix $T_\mathbf{k}$ has the form of
\begin{equation}
	T_\mathbf{k}=\left(\begin{array}{cc}
		\mathbf{u}_{\mathbf{k}} & \mathbf{v}_{2\mathbf{M}-\mathbf{k}}^{*}\\
		\mathbf{v}_\mathbf{k} & \mathbf{u}_{2\mathbf{M}-\mathbf{k}}^{*}
	\end{array}\right)
\end{equation}
where the $3N_L\times 3N_L$ matrices $\mathbf{u}_{\mathbf{k}}$ and $\mathbf{v}_{\mathbf{k}}$ satisfy the normalization condition $\mathbf{u}_{\mathbf{k}}^\dagger \mathbf{u}_{\mathbf{k}}-\mathbf{v}_{\mathbf{k}}^\dagger \mathbf{v}_{\mathbf{k}}=1_{3N_L\times 3N_L}$.

The ground state $|\Omega\rangle$ should be the vacuum of all quasiparticles with $\hat{\gamma}_{j,\mathbf{k}}|\Omega\rangle=0$. Then we consider an excitation $\hat{\gamma}_{j,\mathbf{k}}^{\dagger}|\Omega\rangle$ which is corresponding to the $j$-th column vector of the matrix $T_{\mathbf{k}}$. After one quasiparticle being excited, the particle number changes by an amount
\begin{equation}
	\begin{aligned}
		&\langle\Omega|\hat{\gamma}_{j,\mathbf{k}}\hat{N}\hat{\gamma}_{j,\mathbf{k}}^{\dagger}|\Omega\rangle
		-\langle \Omega|\hat{N}|\Omega\rangle\\
		=&\sum_i(|u_{ij,\mathbf{k}}|^2+|v_{ij,\mathbf{k}}|^2).
	\end{aligned}
\end{equation}
Here, the particle number operator is
\begin{equation}
	\hat{N}=N_0+\sum_{j,{\mathbf{k}}\neq\mathbf{M}}\hat{\beta}_{j,{\mathbf{k}}}^{\dagger}\hat{\beta}_{j,\mathbf{k}}
\end{equation}
with $N_0$ being the particle number of condensate.

We rewrite the $j$-th column vector of the matrix $T_{\mathbf{k}}$ as $|T_{\mathbf{k}}^j\rangle=[\cdots,u_{i\alpha}^j,\cdots,v_{i\alpha}^j,\cdots]^\mathrm{T}$, where $i=1,\cdots,N_L$ denote the original unit cells and $\alpha=x,y,z$. In Fig.~\ref{fig3}(c) of the main text, we show the population of each original unit cell as $\langle \hat{n}_{i,\mathbf{k}}\rangle=\sum_\alpha (|u_{i\alpha}^j|^2+|v_{i\alpha}^j|^2)$ with $\hat{n}_{i,\mathbf{k}}=\sum_\alpha\hat{p}_{i\alpha,\mathbf{k}}^\dagger\hat{p}_{i\alpha,\mathbf{k}}$. The populations of the topological edge states mainly concentrate on the neighborhood of the open boundaries, which illustrates the bulk-boundary correspondence.

\section{Band structure of the 3D cubic optical lattice}
\label{appendixC}
\begin{figure}[htbp]
	\centering
	\includegraphics[width=8.6cm]{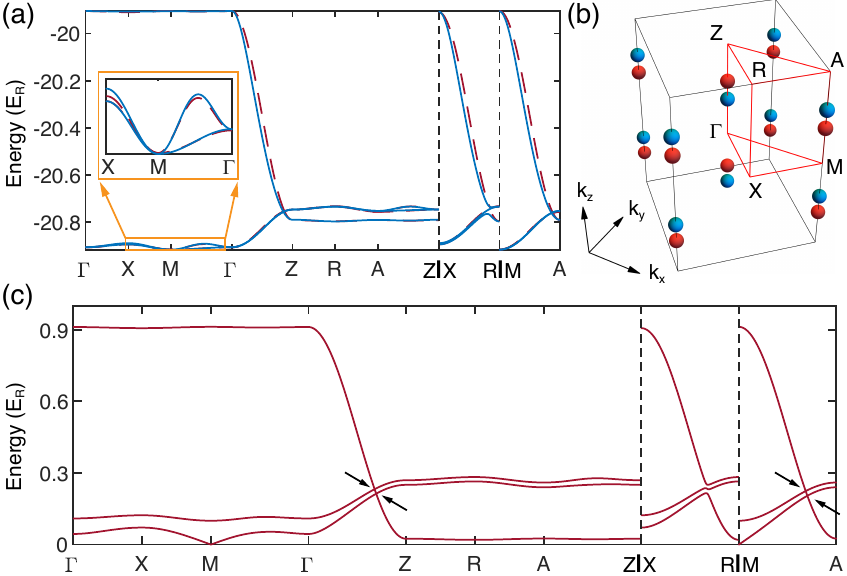}
	\caption{(a) The blue solid lines and red dashed lines denote the single-particle energy spectra obtained by plane wave expansion and the tight-binding model with more hoppings respectively. The inset shows the energy spectra close to the $M$ point. (b) First Brillouin zone of the lattice. The red and blue spheres denote the Weyl points with positive and negative charges respectively. (c) Bogoliubov excitation spectra by applying the tight-binding model with more hoppings. The arrows denote the position of Weyl points.}
	\label{sfig3}
\end{figure}
The 3D cubic optical lattice potential is given by $V(\mathbf{r})=V_\mathrm{2D}(x,y)+V^{\prime}(\mathbf{r})$, where
\begin{equation}
\begin{aligned}
V_\mathrm{2D}(x,y)=&-V_0\big[\cos^2(k_Lx)+\cos^2(k_Ly)\\
&+2\cos(\theta)\cos(k_Lx)\cos(k_Ly)\big]
\end{aligned}
\end{equation}
and
\begin{equation}
	\begin{aligned}
		V^{\prime}(\mathbf{r})=&-V^\prime\cos^{2}[k_{L}(x+z)]-V^\prime\cos^{2}[k_{L}(x-z)]\\
		&-V^\prime\cos^{2}[k_{L}(y+z)]-V^\prime\cos^{2}[k_{L}(y-z)].\\
	\end{aligned}
\end{equation}
The recoil energy $E_\mathrm{R}=\hbar^2k_L^2/(2m_a)$ is used as the unit of the energy, where $m_a$ is the mass of the bosonic atom. We choose lattice parameters as $(V_0,\theta,V^\prime)=(11 E_\mathrm{R},0.484\pi,4.1 E_\mathrm{R})$ and numerically calculate the exact band structure by plane wave expansion. Fig.~\ref{fig3}(a) show the the fourth, fifth, and sixth bands related to $p_{x,y}$ orbitals in deep potential wells and $p_z$ orbitals in shallow potential wells. We find that the fourth and fifth bands are only degenerate at
$\Gamma$ point ($k_x=k_y=k_z=0$) and $M$ point ($k_x=k_L$ and $k_y=k_z=0$) on the plane with $k_z=0$. $M$ point is also the position of the energy minima for the fourth single-particle energy band. Thus, we assume that atoms prefer to condensate at $M$ point in the week interaction limit.

To futher verify the theoretical model in the main text, we consider more hoppings that makes the tight-binding model more realistic. The energy spectra of the tight-binding model with more hoppings are shown in Fig.~\ref{sfig3}(a), which is very close to the exact band structure. The energy minimum is only located at $M$ point. We further use the tight-binding model with more hoppings to study the Bogoliubov excitations. The obtained Bogoliubov excitation spectra and the charges of Weyl points are shown in Fig.~\ref{sfig3}(b, c), which are consistent with the results obtained from the simplified tight-binding model of Eq.~(\ref{tbm}) in the main text.

\bibliography{BibTex}

\end{document}